\documentclass[12pt]{article}

\usepackage{scicite}
\usepackage{graphicx}
\usepackage{wrapfig}
\usepackage[hyphens]{url}
\usepackage{times}
\usepackage{longtable}

\newenvironment{sciabstract}{%
\begin{quote} \bf}
{\end{quote}}

\title{COVID-19 attack rate increases with city size}

\author
{Andrew J. Stier,$^{1\ast}$  Marc G. Berman,$^{1}$
 Luis M. A. Bettencourt$^{2,3}$ \\
\\
\normalsize{$^{1}$Department of Psychology, University of Chicago,}\\
\normalsize{Mansueto Institute for Urban Innovation, University of Chicago}\\
\normalsize{$^{2}$Department of Ecology \& Evolution, University of Chicago}\\
\\
\normalsize{$^\ast$To whom correspondence should be addressed; E-mail:  andrewstier@uchicago.edu.}
}

\date{\today}

\begin{document} 


\baselineskip24pt


\maketitle


\begin{sciabstract}
The current outbreak of novel coronavirus disease 2019 (COVID-19) poses an unprecedented global health and economic threat to interconnected human societies. Until a vaccine is developed, strategies for controlling the outbreak rely on aggressive social distancing. These measures largely  disconnect the social network fabric of human societies, especially in urban areas. Here, we estimate the growth rates and reproductive numbers of COVID-19 in US cities from March 14th through March 19th to reveal a power-law scaling relationship to city population size. This means that COVID-19 is spreading faster on average in larger cities with the additional implication that, in an uncontrolled outbreak, larger fractions of the population are expected to become infected in more populous urban areas. We discuss the implications of these observations for controlling the COVID-19 outbreak, emphasizing the need to implement more aggressive distancing policies in larger cities while also preserving socioeconomic activity.
\end{sciabstract}


The coronavirus pandemic of 2019-20 (COVID-19) is an unprecedented worldwide event. Its speed of propagation, its international reach and the unprecedented coordinated measures for its mitigation, are only possible in a world that is more connected and more urbanized than at any other time in history.

As a novel infectious disease in human populations, COVID-19 has a number of quantitative signatures to its pattern of spread. These signatures make its dynamics more difficult to contain but also easier to understand.  

First, because there is no history of previous exposure, all human populations in contact with the virus are (presumably) susceptible. This means that the susceptible population is the world's total population writ large. Second, because COVID-19 is a respiratory disease, it is easily transmissible resulting in high reproductive numbers, $R=2.2-6.5$ \cite{reproductiveNumberMeta,lancetRo,yang2020early}, though considerable uncertainty remains about these estimates. Third, COVID-19 appears to be characterized by reproductive numbers above the epidemic threshold ($R>1$) everywhere around the world, regardless of environmental conditions such as humidity or temperature. These reproductive numbers are considerably higher than seasonal influenza \cite{biggerstaff2014estimates}. Bringing the disease reproductive number below the epidemic threshold ($R \rightarrow R<1$) is the main goal of all public health interventions; once this is achieved the disease's transmission chain reaction will shut down. 

The reproductive number is the product of two factors $R=\beta/\gamma$, the infectious period $1/\gamma$ (a physiological property) and the contact rate $\beta$, which is a property of the population, essentially measuring the number of social contacts that can transmit the disease per unit time. Of these, only the contact rate can be changed via public health interventions. 

In the absence of a vaccine, social distancing remains the only option to slow down the spread of the disease and arrest potential mortality. Governments around the world are now enacting aggressive policies, including "shelter in place" and emergency closures of all non-essential services, which carry severe economic and social consequences. For example, in the last week, the US Centers for Disease Control and the White House have recommended extreme social distancing in order to slow down the current outbreak of novel coronavirus disease 2019 (COVID-19) \cite{whitehouse}. However, these measures are less aggressive than what has been put in place elsewhere \cite{chinaNewYear}. There is still a great deal of uncertainty as to how strong social distancing recommendations must be or how long they must last. Importantly, national and regional social distancing policies are likely to impact individual cities differently.

Cities are predicated on extensive and intense socioeconomic interactions. Many of their measurable properties - from the size of their economies, to their crime rates, to the prevalence of certain infectious diseases - are mediated by socioeconomic interactions. These interactions are subject to well known scaling effects, which are magnified by city population size \cite{luisScience}. All of these relationships are tied to socioeconomic networks with average degree (number of social connections per capita) that increases approximately as a power law of city size $k(N)=k_0 N^\delta$, with $\delta\simeq 1/6$~ \cite{luisScience,schlapfer2014scaling}. 

This is a large effect. Based on data of mobile phone social networks \cite{schlapfer2014scaling}, people living in a city of 500,000 have, on average, 11 people in their mobile phone social network, while people living in a city of 5,000,000 people have approximately 15.  This is relevant to disease transmission as the average contact rate is proportional to degree $\beta \propto k(N)$ (see Materials and Methods). Therefore, we expect that initial growth rates of COVID-19 cases to be higher in larger cities (see Materials and Methods). This is what is found empirically (see Figure 1A).  
\begin{figure}[ht]
    \centering
    \includegraphics[width=1\textwidth,trim={.8cm 0 1.2cm 0cm},clip]{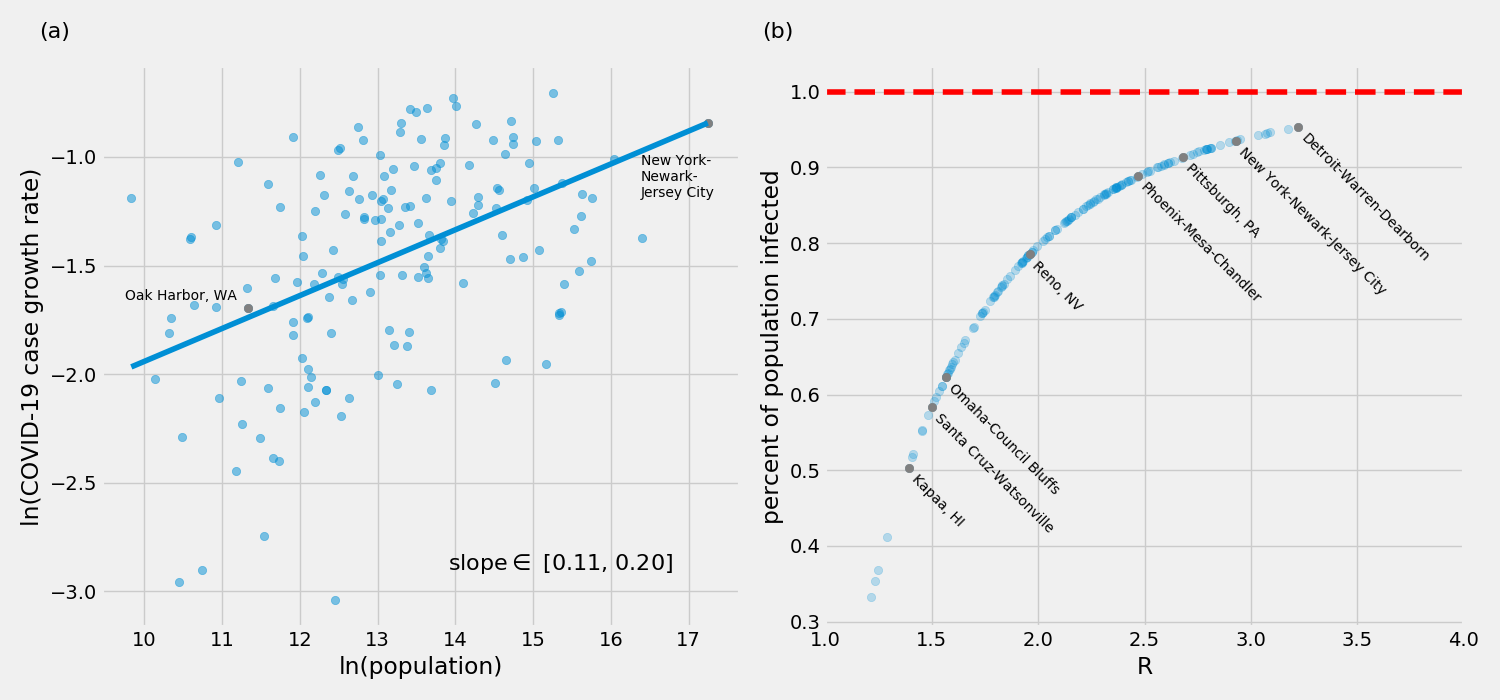}
    \caption{COVID-19 reported cases grow faster in larger cities. (a) Estimated exponential daily growth rates of COVID-19 in US Metropolitan Areas (MSAs). These estimates were made with the assumption that cities were experiencing exponential growth of cases. The growth rate of COVID-19 cases is approximately 2.4 times faster in New York-Newark-Jersey compared to Oak Harbour, WA (b) In the absence of effective controls, larger cities are expected to have more extensive epidemics than smaller cities, Eq. (1). Higher values of $R$ result in a greater percentage of the population eventually infected, unless this effect is curbed by controls that reduce the social contact rate. The translation of growth rates into reproductive numbers was obtained using an infectious period of $1/\gamma=4.5$ days. These estimated values of $R$ are high in some cases (e.g. New York City) compared to reports in other situations and may in part be the result of the acceleration of testing in larger cities and specific places.}
\end{figure} 

 A larger reproductive number for spreading processes in larger cities has two important consequences \cite{dalziel2018urbanization,bettencourt2007invention,feldman1999innovation,ACS2004635}. First, the reproductive number, $R$, sets a finite threshold for how an epidemic outbreak propagates in a population, just like the branching rate in a chain reaction \cite{anderson1992infectious,bettencourt2006power}. For $R<1$, an introduced disease will die off because it will be dampened in transmission, while for $R>1$, disease transmission will be amplified and result in an epidemic where the disease is transmitted quickly to almost everyone (see Figure 1B). Because we expect $R\sim N^\delta$ to increase with city size, we expect larger cities to be more susceptible to both contagious diseases, but also to the spread of information (see below).   

Second, the size of an epidemic outbreak, as measured by the percent of the population that becomes infected, is also related to the reproductive number. In complex epidemic models, this needs to be computed numerically, but for a simple Susceptible-Infected-Recovered (SIR) model \cite{anderson1992infectious} we can integrate the dynamics and write the explicit expression (see Methods)
\begin{equation}
\frac{S_\infty}{S_0}=e^{- R (1-S_\infty/N)} 
\end{equation}
where $S_0$ is the initial suscepible population size (before the outbreak) and $S_\infty$ is the (smaller) final population of susceptible people. A larger $R\sim~N^\delta$ leads to more extensive epidemics. The percent of people infected at the end of the outbreak is $1-S_\infty/N$  which is larger in populations with larger $R$, such as in larger cities.   A final point centers on the fraction of the population that must be removed from the susceptible class when $R>1$ to stop the outbreak. This is often called the vaccination rate, $p_R$, which is equally relevant in the context of social distancing (as only the means of the intervention and its duration differ). In the SIR model, this is simply $p_R=1-1/R$, which shows that as cities get larger the distancing rate must also increase (see Supplementary Figure 1). 

These observations have a number of implications that can inform evolving national, regional, and local responses to the outbreak of COVID-19. First, it is particularly important for larger cities to act quickly to contain this outbreak. Second, social distancing will impact cities differently based on city size. From the perspective of containing the outbreak, larger cities require more aggressive social distancing policies, corresponding to a larger $p_R$. At the same time, once the outbreak is contained it might be possible to relax social distancing policies in smaller cities first, allowing a faster return to normal life and economic activity compared to more populous urban areas. 

These distinctions may help to bring more nuance to ongoing strategies for suppression and control of COVID-19, including gradually restoring socioeconomic activity in context appropriate and safe ways.

Because of their higher network density, insufficient social distancing in larger or typically more connected cities may lead to bigger outbreaks and to the creation of reservoirs for the disease, which can continue to create introductions elsewhere. These dynamics may also play out within cities, as communities in which people interact more densely from the perspective of disease transmission (e.g., downtowns), may similarly act as contagion reservoirs that may prolong the duration of the present outbreak and potentially create secondary reinfection waves.

Finally, as strategies for controlling this outbreak continue to evolve, it is critical to keep in mind that almost everything that we appreciate about urban environments, including their economic prowess, their ability to innovate, and their role in their inhabitants social and mental health, is predicated on network effects mediated by socioeconomic interactions. The ability to succeed against a fast emerging epidemic like COVID-19 depends on preserving as much person-to-person connectivity (e.g., through technology), while stopping disease transmission. Research on safe types of socioeconomic contact and exchange is therefore paramount so that we can succeed in controlling this outbreak while maintaining livelihoods, socioeconomic capacity, and mental health. This can in principle be done through approaches that make the most of emerging, real-time data to create context appropriate suppression strategies at local, regional, national, and global levels.

The higher socioeconomic connectivity of larger cities in a fast urbanizing world makes containing emergent epidemics harder. But the density of socioeconomic connections in cities can also facilitate the spread of information, social coordination, and innovations necessary to stop the spread of COVID-19. This information and associated actions can easily spread much faster than the biological viral contagion. To fight an exponential,  we need to create an even faster exponential!


\bibliography{scibib}

\bibliographystyle{Science}

\section*{Acknowledgments}
This research is partially supported by the Mansueto Institute for Urban Innovation and a Social Science Research grant from the University of Chicago.

\section*{Supplementary materials}

\section*{Materials and Methods}
\noindent{\bf Data and Urban Units of Analysis:}
 Here we briefly describe the mathematical analysis steps. County level daily data from March 13-24 were aggregated to the city level (Metropolitan Statistical Areas, which are integrated labor markets) using delineation files from the US Office of Budget and Management \cite{omb}. Counties with at least 1 reported case of COVID-19 are available in the data. We next excluded cities that had no COVID-19 cases on March 13. This excluded cities with low case counts, likely due to introductions from outside the MSA, for which accurate estimates of local case growth rate is unlikely. Results were similar for all contiguous subsets of the data of at least 7 days (see Supplmentary Figure 4a). This left 163 cities for further analysis. We substracted deaths from cases in each city and found the slope of the $ln(cases)\sim ln(a) +r\cdot t$ line for each resulting time-series of active COVID-19 cases. Finally we plotted the natural logarithm of $r$ and the natural logarithm of city population from 2018 census estimates \cite{census}, and performed an ordinary least squares linear regression to determine the slope of the scaling line. Regression residuals were not related to city population (Supplementary Figure 2) and a q-q plot of the residuals indicated that they were well described by a normal distribution (Supplementary Figure 3).
 
 In order to estimate the reproductive number $R$ we multiplied the growth rate of each city, $r$, by an average infections period of $1/\gamma=4.5$ days and adding one (see below). The size of the epidemic was then estimated by finding the root of the equation $y=\ln(x)+R\cdot (1-x)$, where $x = S_\infty/N$.
 
 Accurately estimating the growth rate of epidemics is often difficult \cite{Yadlowsky2020}, however, here we are concerned with the pattern of growth rates among cities rather than the precision of our growth rate estimates. To that end we additionally estimated the growth rate of COVID-19 cases by $r=ln(cases_T/cases_0)/T$ which is an estimate of the slope of the line $ln(cases)\sim ln(a) +r\cdot t$ from the first and last points of the time series (Supplementary Figure 4b). These growth rate estimates showed a scaling relationship with city size that is consistent with Figure 1 of the main text. This was observed despite variations in growth rate estimates between the two methods

 \noindent{\bf Epidemic Models and the Reproductive Rate}
Even though well known, we include here the basic derivation of the reproductive rate and final size of the epidemic epidemic models, for the sake of completeness.

\begin{figure}[ht]
    \centering
    \includegraphics[width=0.8\textwidth,trim={.8cm 0 1.2cm 0cm},clip]{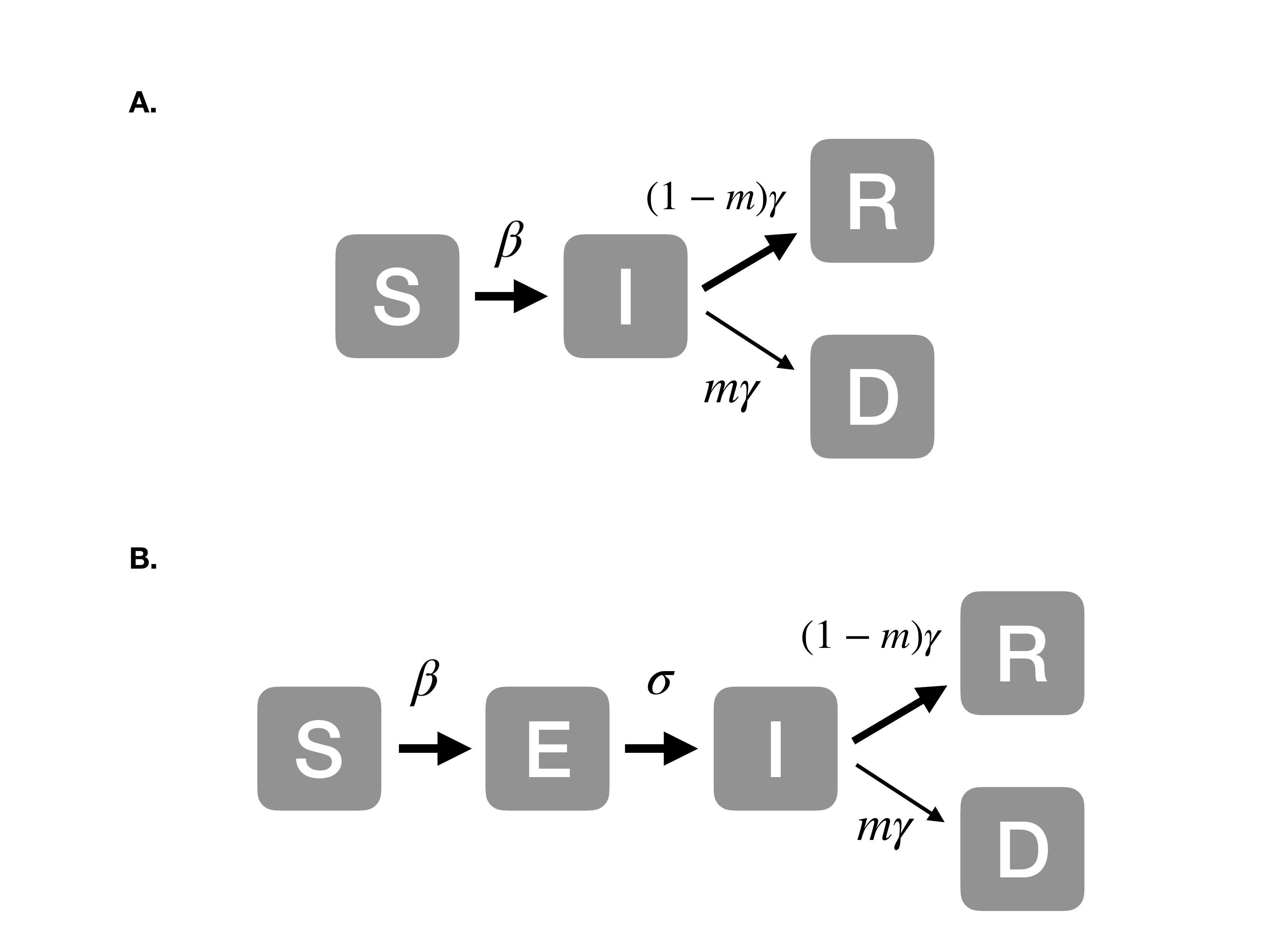}
    \caption{ {\bf Epidemic models and their parameters.}Epidemic models divide a  population of size $N$, into subgroups by their health status at each time: susceptible (S), exposed (E), infectious (I), recovered (R) and dead (D), such that $N=S+E+I+R+D$. Arrows show temporal progression(see text for the mathematical implementation of the diagram.) A. The simpler SIR  model. B. The SEIR model, which includes in addition to  the SIR model a class of exposed individuals, which are {\it not} infectious.}
\end{figure}

The SIR model (Figure 2A) is the simplest relevant description of an epidemic in a population. The model is typically written in terms differential equations as
\begin{eqnarray}
\frac{d S}{dt}= - \beta \frac{S}{N} I, \quad \frac{d I}{dt}=  \beta \frac{S}{N} I - \gamma I, \quad \frac{d R}{dt}= (1-m) \gamma I, \quad \frac{d D}{dt}= m \gamma I.
\end{eqnarray}
Here, $\beta$ is known as the contact rate, $1/\gamma$ is the infectious period, $1/\sigma$ is the non-infectious incubation time, and $m$ is the probability that an infected individual dies (mortality rate). 

The reproductive rate $R$ can be easily deduced from the initial growth  of $I$, when $S/N\simeq 1$, which is 
\begin{eqnarray}
\frac{d I}{dt}=  \gamma \left[ R \frac{S}{N} - 1 \right]I,
\end{eqnarray}
with $R=\beta/\gamma$. We see that the temporal growth rate $r=\gamma \left[ R \frac{S}{N} - 1 \right]\simeq \gamma (R-1)$. We see that the number of cases will grow exponentially if $R>1$ and, conversely, decay exponentially when $R<1$.

At a later time, as the susceptible population becomes depleted, the effective $R$ decreases. Distancing or vaccination work by removing people from the susceptible class, which can be modelled by reducing $S/N$ or, equivalently  for $R$, $\beta$ since these two factors always appear multiplied together i. 
the y

To obtain the expression for the final size of the epidemic we note that
\begin{equation}
\int_0^\infty dt' \frac{d (S+I)}{dt'} = S_\infty - S_0 - I_0 = - \gamma \int_0^\infty dt' I(t'), 
\end{equation}
where $S_\infty$ is the population of susceptibles left uninfected at the end of the outbreak, and $S_0, \ I_0$ are the size of the susceptible and infected population at the initial time.

The equation for $S$ in time can be written as
\begin{equation}
    \frac{d S}{dt}= - \beta \frac{S}{N} I \rightarrow \frac{d \ln S}{dt} = - \frac{\beta}{N}  I,
\end{equation} 
which now can be integrated to give
\begin{equation}
    \ln \frac{S_\infty}{S_0}= - R (1 - \frac{S_\infty}{N} ),
\end{equation}
which is the  desired result, used in the main text to create Figure 1B.

Finally we note that even though a model with a non-infectious incubation period such as the SEIR looks more complex, it has the same value of $R$ provided we can neglect mortality in the $S, E$ classes, not due to the disease outbreak.

\noindent{\bf Derivation of the City Size Dependence of the Reproductive Number}

The most important quantity characterizing epidemic processes is the reproductive number, $R$, which measures the number of secondary cases induced by an infectious individual in a fully susceptible population. Recall that we expect that human population are thought to be wholly susceptible to COVID-19.

For a contagion network, the reproductive number is related to the statistics of degree, $k$, as
\begin{equation}
    R= p_I (\frac{\langle k^2 \rangle}{\langle k \rangle})=p_I \langle k \rangle \left(1-\frac{\sigma_k^2}{\langle k \rangle^2}.    \right),                                       
\end{equation}
where $p_I$  is the infection probability per contact, $\langle ... \rangle$ denotes expectation values over the population and $\sigma_k^2$ is the degree variance. 

For a lognormal degree distribution, which is typical of social networks in cities, the degree average and variance are given by 
\begin{equation}
    \langle k \rangle =e^{\mu+\sigma^2/2},   \quad     \sigma_k^2=\left( e^{\sigma^2 }-1 \right) e^{2 \mu +\sigma^2 }, 
\end{equation}
with the parameters $\mu = \langle \ln k \rangle $, $ \sigma^2= \langle \left(\ln k -\angle \ln k \rangle \right)^2 \rangle$. This results in a simple and elegant expression for the reproductive number
\begin{equation}
R (N)=p_I e^{\sigma^2}  \langle k(N) \rangle = p_I k_0  e^{\sigma^2} N^\delta,
\end{equation}
where, in the last equality, we introduced the scaling relation for the average degree with city population size, $\langle k(N) \rangle =k_0 N^\delta$. We see therefore that in general the reproductive number is expected to be a function of city size $N$, and to be larger in bigger cities. How much bigger, depends on the behavior of the log-variance, $\sigma^2$, and whether this parameter is city size dependent, an issue that can generates statistical corrections beyond “mean-field” to the simplest expectations from urban scaling theory with $\delta \simeq 1/6$.

\newpage
\clearpage
\section*{Supplementary Figures}
\setcounter{figure}{0}
\renewcommand{\figurename}{Supplementary Figure}
\begin{figure}[h]
    \centering
    \includegraphics[width=\textwidth,trim={0 0 0 0},clip]{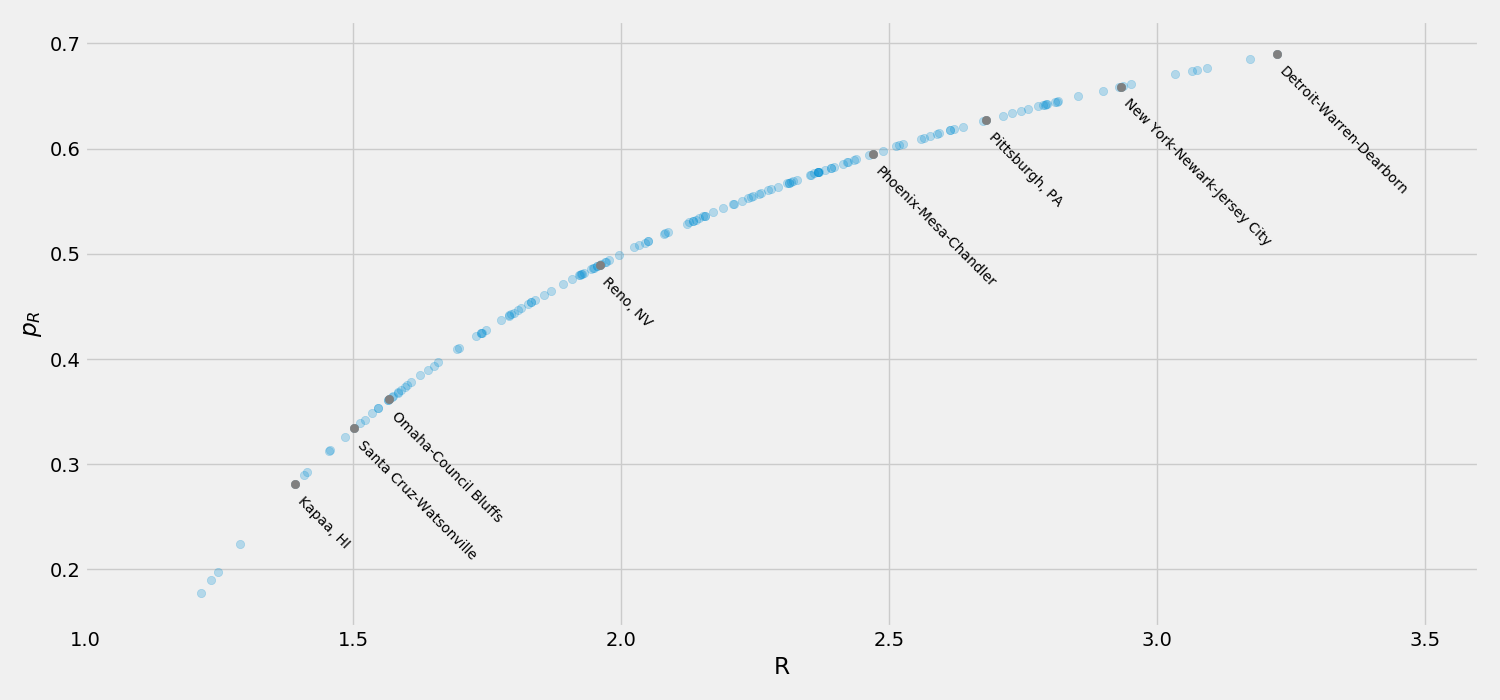}
    \caption{Larger cities are expected to require more aggressive strategies to control epidemics than smaller cities. Higher values of $R$ result in a a greater required vaccination rate, $p_R=1-1/R$, in order to stop the outbreak. This rate applies to social distancing as well as vaccination and herd immunity. The translation of growth rates into reproductive numbers was obtained using an infectious period of $1/\gamma=4.5$ days. These estimated values of $R$ are high in some cases (e.g. New York City) compared to reports in other situations and may in part be the result of the acceleration of testing in larger cities and specific places.}
    \label{fig:S1}
\end{figure}

\begin{figure}[h]
    \centering
    \includegraphics[width=\textwidth,trim={0 0 0 0},clip]{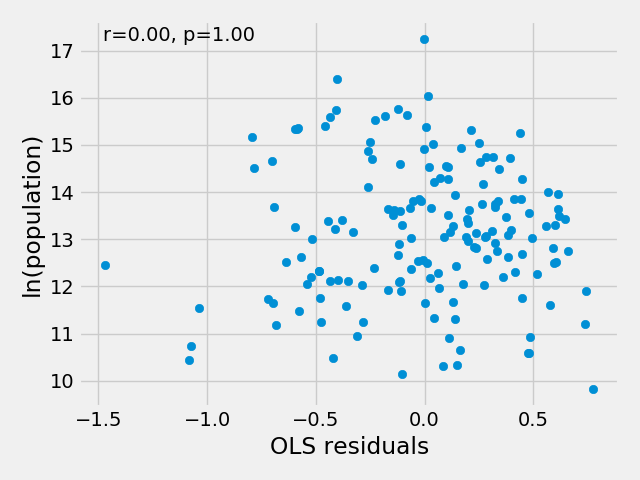}
    \caption{Scaling residuals are unrelated to city size. The lack of a correlation between the residuals of the scaling fit (Figure 1a) and the logarithm population indicate that a correction to the scaling exponent is not necessary. Residuals were calculated from the OLS fit to the plot of logarithm of COVID-19 case growth rates and the logarithm of population, where growth rates were estimated with linear regression to the COVID-19 case time-series.}
    \label{fig:S2}
\end{figure}

\begin{figure}[h]
    \centering
    \includegraphics[width=\textwidth,trim={0 0 0 0},clip]{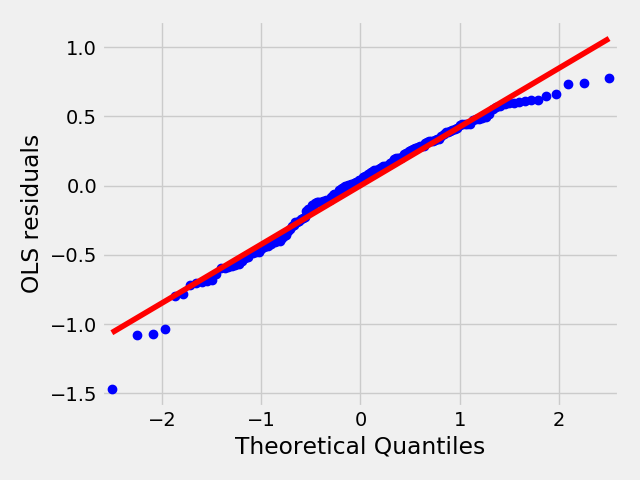}
    \caption{Scaling residuals roughly normal distributed. Residual from the scaling fit (Figure 1a) roughly match theoretical quantiles of a normal distribution, indicating a good linear fit. Residuals were calculated from the OLS fit to the plot of logarithm of COVID-19 case growth rates and the logarithm of population, where growth rates were estimated with linear regression to the COVID-19 case time-series. }
    \label{fig:S3}
\end{figure}

\begin{figure}[h]
    \centering
    \includegraphics[width=\textwidth,trim={0 0 0 0},clip]{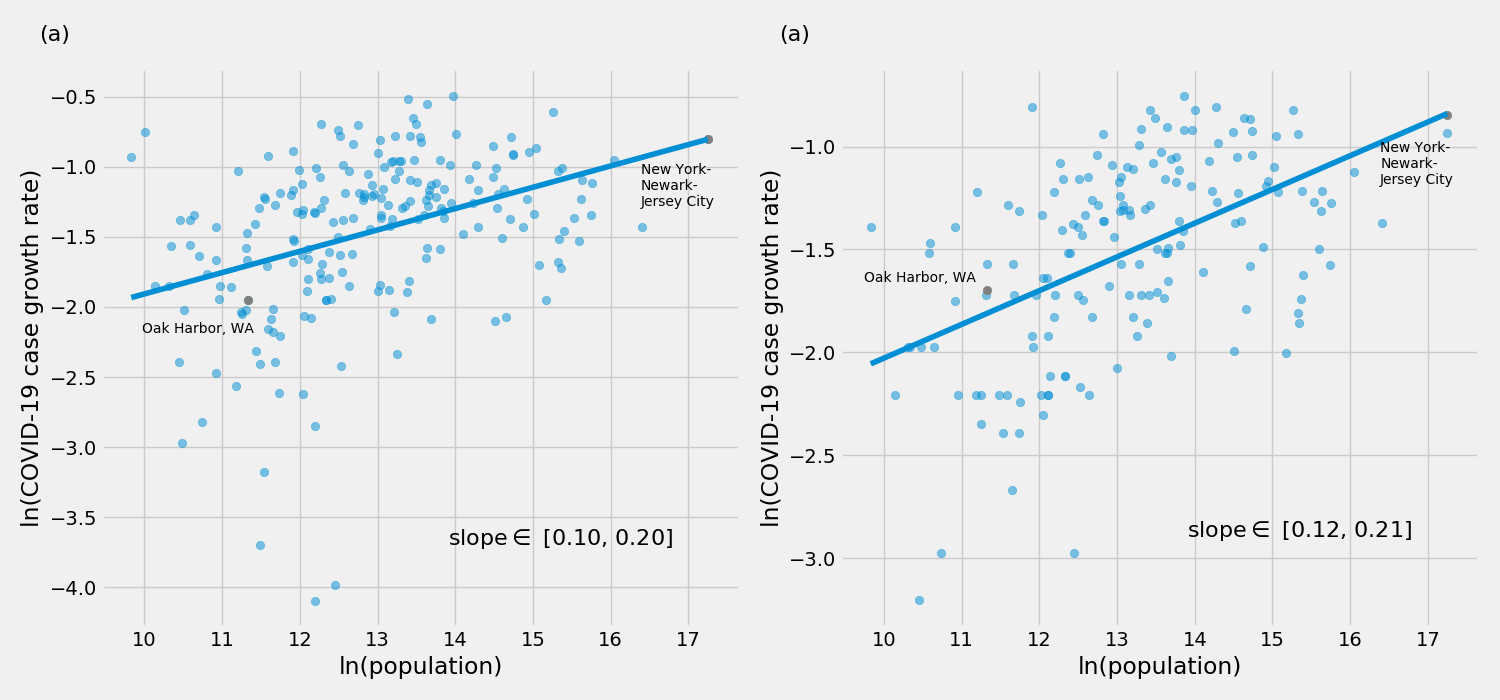}
    \caption{Scaling Plots for alternate methods for estimating COVID-19 case growth rates. Both alternate methods produce growth rates which show scaling relationships with city size consistent with Figure 1a. Growth rates were estimated with (a) a linear regression, as in Figure 1a, to data from March 16th to March 23rd (b) $r=ln(cases_T/cases_0)/T$ to data from March 13th to March 25th. This is an estimate of the slope of the line $ln(cases)\sim ln(a) +r\cdot t$ from the first and last points of the time series.}
    \label{fig:S4}
\end{figure}

\begin{figure}[h]
    \centering
    \includegraphics[width=\textwidth,trim={0 0 0 0},clip]{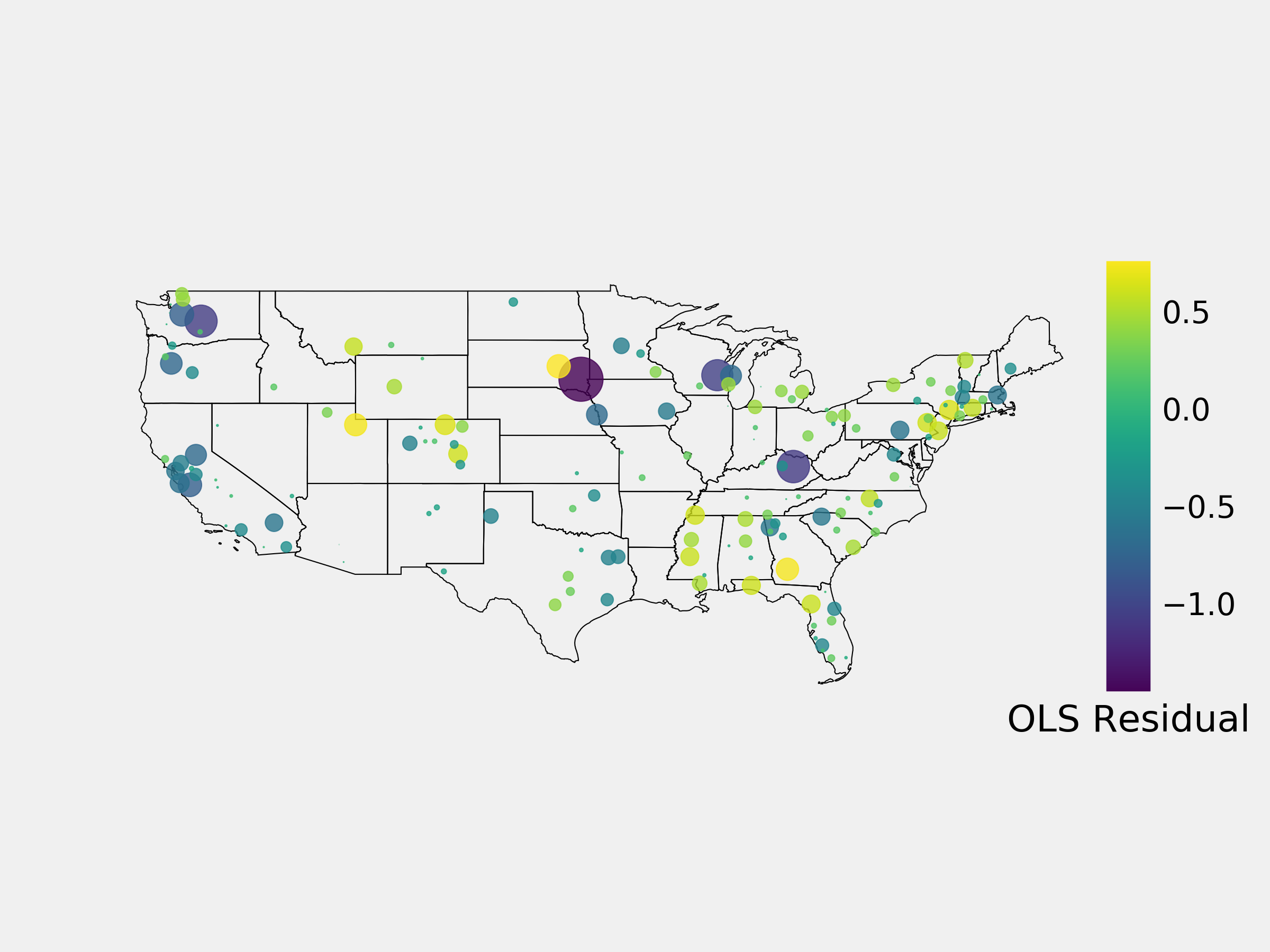}
    \caption{Map of residuals of the scaling fit in Figure 1a}
    \label{fig:S5}
\end{figure}

\clearpage
\section*{Supplementary Tables}
\textbf{Supplemetary Table 1: Estimated COVID-19 daily growth rates in MSA}\\
\begin{longtable}{ll}
\textit{Estimated daily COVID-19} \\\textit{Case Growth Rate} & \textit{MSA} \\

                                                         0.43 &           New York-Newark-Jersey City, NY-NJ-PA \\
                               0.25 &              Los Angeles-Long Beach-Anaheim, CA \\
                               0.36 &              Chicago-Naperville-Elgin, IL-IN-WI \\
                               0.30 &                 Dallas-Fort Worth-Arlington, TX \\
                               0.23 &            Houston-The Woodlands-Sugar Land, TX \\
                               0.31 &         Miami-Fort Lauderdale-Pompano Beach, FL \\
                               0.28 &     Philadelphia-Camden-Wilmington, PA-NJ-DE-MD \\
                               0.22 &    Washington-Arlington-Alexandria, DC-VA-MD-WV \\
                               0.26 &            Atlanta-Sandy Springs-Alpharetta, GA \\
                               0.21 &                                    Longview, TX \\
                               0.33 &                       Phoenix-Mesa-Chandler, AZ \\
                               0.18 &              San Francisco-Oakland-Berkeley, CA \\
                               0.18 &            Riverside-San Bernardino-Ontario, CA \\
                               0.18 &                  Boston-Cambridge-Newton, MA-NH \\
                               0.40 &                               Oklahoma City, OK \\
                               0.49 &                     Detroit-Warren-Dearborn, MI \\
                               0.14 &                     Seattle-Tacoma-Bellevue, WA \\
                               0.24 &         Minneapolis-St. Paul-Bloomington, MN-WI \\
                               0.40 &                                St. Louis, MO-IL \\
                               0.32 &              San Diego-Chula Vista-Carlsbad, CA \\
                               0.36 &             Tampa-St. Petersburg-Clearwater, FL \\
                               0.30 &                   Baltimore-Columbia-Towson, MD \\
                               0.23 &                      Denver-Aurora-Lakewood, CO \\
                               0.40 &               Charlotte-Concord-Gastonia, NC-SC \\
                               0.39 &                   Orlando-Kissimmee-Sanford, FL \\
                               0.43 &                   San Antonio-New Braunfels, TX \\
                               0.23 &             Portland-Vancouver-Hillsboro, OR-WA \\
                               0.14 &                 Sacramento-Roseville-Folsom, CA \\
                               0.37 &                                  Pittsburgh, PA \\
                               0.26 &                Las Vegas-Henderson-Paradise, NV \\
                               0.32 &                              Kansas City, MO-KS \\
                               0.32 &                            Cleveland-Elyria, OH \\
                               0.29 &                Indianapolis-Carmel-Anderson, IN \\
                               0.13 &              San Jose-Sunnyvale-Santa Clara, CA \\
                               0.40 &                                    Columbus, OH \\
                               0.30 &                       Providence-Warwick, RI-MA \\
                               0.31 &  Nashville-Davidson--Murfreesboro--Franklin, TN \\
                               0.43 &                          Milwaukee-Waukesha, WI \\
                               0.28 &                                Jacksonville, FL \\
                               0.35 &                Austin-Round Rock-Georgetown, TX \\
                               0.21 &                                Raleigh-Cary, NC \\
                               0.46 &           Hartford-East Hartford-Middletown, CT \\
                               0.48 &                               Memphis, TN-MS-AR \\
                               0.30 &              Louisville/Jefferson County, KY-IN \\
                               0.40 &                                   Rochester, NY \\
                               0.39 &                           Birmingham-Hoover, AL \\
                               0.25 &                                      Tucson, AZ \\
                               0.25 &                       Grand Rapids-Kentwood, MI \\
                               0.24 &                                      Fresno, CA \\
                               0.36 &                              Urban Honolulu, HI \\
                               0.35 &                 Bridgeport-Stamford-Norwalk, CT \\
                               0.33 &                                Worcester, MA-CT \\
                               0.35 &                     Albany-Schenectady-Troy, NY \\
                               0.13 &                     Omaha-Council Bluffs, NE-IA \\
                               0.26 &                                   Knoxville, TN \\
                               0.23 &                Oxnard-Thousand Oaks-Ventura, CA \\
                               0.21 &                                     El Paso, TX \\
                               0.46 &               Allentown-Bethlehem-Easton, PA-NJ \\
                               0.30 &                                    Columbia, SC \\
                               0.22 &                                 Albuquerque, NM \\
                               0.22 &               North Port-Sarasota-Bradenton, FL \\
                               0.40 &                 Charleston-North Charleston, SC \\
                               0.21 &                                    Stockton, CA \\
                               0.27 &                       Cape Coral-Fort Myers, FL \\
                               0.45 &                            Colorado Springs, CO \\
                               0.35 &                                       Akron, OH \\
                               0.29 &                                  Boise City, ID \\
                               0.46 &            Poughkeepsie-Newburgh-Middletown, NY \\
                               0.16 &                                       Tulsa, OK \\
                               0.15 &          Deltona-Daytona Beach-Ormond Beach, FL \\
                               0.29 &                                     Madison, WI \\
                               0.21 &                                     Wichita, KS \\
                               0.43 &                                     Jackson, MS \\
                               0.41 &                          Durham-Chapel Hill, NC \\
                               0.27 &                               Winston-Salem, NC \\
                               0.13 &                         Harrisburg-Carlisle, PA \\
                               0.15 &                                     Modesto, CA \\
                               0.35 &               Youngstown-Warren-Boardman, OH-PA \\
                               0.32 &                              Chattanooga, TN-GA \\
                               0.26 &                                Fayetteville, NC \\
                               0.17 &                           Lexington-Fayette, KY \\
                               0.29 &                         Santa Rosa-Petaluma, CA \\
                               0.34 &                        Lansing-East Lansing, MI \\
                               0.30 &                                    Richmond, VA \\
                               0.30 &   Myrtle Beach-Conway-North Myrtle Beach, SC-NC \\
                               0.25 &                                     Visalia, CA \\
                               0.28 &                                 Springfield, MO \\
                               0.21 &                                        Reno, NV \\
                               0.37 &                                  Huntsville, AL \\
                               0.14 &                                     Vallejo, CA \\
                               0.27 &                                       Salem, OR \\
                               0.31 &                            Ogden-Clearfield, UT \\
                               0.20 &                            Canton-Massillon, OH \\
                               0.28 &                         Naples-Marco Island, FL \\
                               0.28 &                                   Ann Arbor, MI \\
                               0.40 &                           Trenton-Princeton, NJ \\
                               0.30 &                              Killeen-Temple, TX \\
                               0.42 &                                Fort Collins, CO \\
                               0.34 &                     South Bend-Mishawaka, IN-MI \\
                               0.19 &                                  Montgomery, AL \\
                               0.12 &                                 Spartanburg, SC \\
                               0.31 &                                     Greeley, CO \\
                               0.28 &                                  Utica-Rome, NY \\
                               0.21 &      Virginia Beach-Norfolk-Newport News, VA-NC \\
                               0.20 &                      Olympia-Lacey-Tumwater, WA \\
                               0.11 &                      Santa Cruz-Watsonville, CA \\
                               0.38 &          Crestview-Fort Walton Beach-Destin, FL \\
                               0.38 &                                 Gainesville, FL \\
                               0.21 &           Bremerton-Silverdale-Port Orchard, WA \\
                               0.05 &                                 Sioux Falls, SD \\
                               0.24 &                                      Yakima, WA \\
                               0.16 &                                  Binghamton, NY \\
                               0.19 &                                  Tuscaloosa, AL \\
                               0.13 &                                    Hereford, TX \\
                               0.13 &                                       Tyler, TX \\
                               0.31 &                                  Bellingham, WA \\
                               0.22 &                                  Lebanon, NH-VT \\
                               0.34 &                 Burlington-South Burlington, VT \\
                               0.12 &                                   St. Cloud, MN \\
                               0.29 &                                   Rochester, MN \\
                               0.20 &                                      Racine, WI \\
                               0.13 &                                        Bend, OR \\
                               0.18 &                                  Torrington, CT \\
                               0.14 &                                   El Centro, CA \\
                               0.13 &                                 Punta Gorda, FL \\
                               0.18 &                                    Kingston, NY \\
                               0.11 &                                   Iowa City, IA \\
                               0.23 &                                    Billings, MT \\
                               0.26 &                            East Stroudsburg, PA \\
                               0.15 &                                      Pueblo, CO \\
                               0.21 &                                      Madera, CA \\
                               0.16 &                                    Santa Fe, NM \\
                               0.17 &                                 Hattiesburg, MS \\
                               0.40 &                                      Albany, GA \\
                               0.12 &                                  Pittsfield, MA \\
                               0.29 &                      Mount Vernon-Anacortes, WA \\
                               0.09 &                              Albany-Lebanon, OR \\
                               0.21 &                                  Morristown, TN \\
                               0.18 &                                    Valdosta, GA \\
                               0.09 &                                   Sheboygan, WI \\
                               0.33 &                                     Bozeman, MT \\
                               0.13 &                             Lewiston-Auburn, ME \\
                               0.06 &                                 Fond du Lac, WI \\
                               0.10 &                                        Rome, GA \\
                               0.18 &                                  Oak Harbor, WA \\
                               0.20 &                                      Kokomo, IN \\
                               0.11 &                            Glenwood Springs, CO \\
                               0.13 &                                       Minot, ND \\
                               0.36 &                                       Heber, UT \\
                               0.09 &                                       Kapaa, HI \\
                               0.12 &                                     Calhoun, GA \\
                               0.27 &                                    Picayune, MS \\
                               0.18 &                                     Edwards, CO \\
                               0.05 &                                  Ellensburg, WA \\
                               0.19 &                                   Cedartown, GA \\
                               0.25 &                                    Riverton, WY \\
                               0.25 &                                   Greenwood, MS \\
                               0.10 &                                  Bennington, VT \\
                               0.05 &                              Mount Sterling, KY \\
                               0.18 &                                Breckenridge, CO \\
                               0.16 &                                    Sheridan, WY \\
                               0.13 &                           Steamboat Springs, CO \\
                               0.30 &                                       Huron, SD \\
\end{longtable}

\bibliography{scibib}

\bibliographystyle{Science}
\end{document}